\documentclass{iopconfser}
\usepackage[utf8]{inputenc}
\usepackage{graphicx}
\usepackage{float}
\usepackage{amssymb}
\usepackage{amsmath}  
\usepackage{adjustbox}
\usepackage{enumitem}
\usepackage{mathtools}
\usepackage{dsfont}
\usepackage{array}
\usepackage{mathrsfs}
\usepackage{multirow}
\usepackage{upgreek}
\usepackage{xcolor}
\usepackage{bm}
\usepackage{bbm}
\usepackage{cite}
\usepackage[colorlinks, citecolor=blue, urlcolor=blue]{hyperref}
\begin{document}

\title{Quantum-centric Supercomputing for Physics Research}

\author{Vincent R.\ Pascuzzi, Antonio D. C\'{o}rcoles}

\affil{IBM Quantum, IBM Thomas J.\ Watson Research Center, USA
}

\email{vrpascuzzi@ibm.com, adcorcol@us.ibm.com}

\begin{abstract}
    This document summarizes the presentation on Quantum-centric Supercomputing~\cite{pres} given at the 22$^{\text{nd}}$ International Workshop on Advanced Computing and Analysis Techniques in Physics Research, hosted at Stony Brook University.
\end{abstract}

\section{Introduction\label{sec:intro}}
The rise of parallel computing, in particular graphics processing units (GPU), and machine learning and artificial intelligence has led to unprecedented computational power and analysis techniques.
Such technologies have been especially fruitful for theoretical and experimental physics research where the embarrassingly parallel nature of certain workloads -- e.g., Monte Carlo (MC) event generation, detector simulations, workflows, and data analysis -- are exploited to attain significant performance improvements.
Despite these capabilities, there still exist an array of problems that are manifestly intractable with classical computation alone, or for which classical computation provides only approximate or inefficient solutions.

Quantum computing is able to give exponential gains in both time and space for certain classes of problems.
However, quantum workloads require significant classical computing support for preprocessing, including optimization and compilation, and postprocessing.
This naturally leads to the concept of Quantum-centric Supercomputing (QCSC):
the integration of quantum and classical computational resources enabling the execution of parallel and asynchronous hybrid workloads.
For example, HPC-assisted quantum computation can help to extract or boost useful signals in utility-scale experiments~\cite{robledomoreno2024chemistryexactsolutionsquantumcentric}.
IBM Quantum is engaging with the scientific and HPC communities to deliver them unrivaled quantum computing capabilities that will play a central role in the most powerful supercomputing systems in the world.

In this talk, we gave a brief overview of IBM Quantum's development roadmap and showed how QCSC naturally fits this vision.
We explored ways in which the physics community and computational scientists can benefit from QCSC, and detailed a use case in high energy physics (HEP) suitable for these integrated systems.

\section{Complex computations and the IBM Quantum Roadmap\label{sec:complex}}
High performance computing systems are undoubtedly powerful.
However, as a result of the quantum Hilbert space, simulating quantum systems classically scales exponentially in both time and space as circuit complexity increases.
Before quantum error correction (QEC) and fault-tolerance, we have observed “quantum utility”~\cite{Kim2023}: the demonstration of a noisy quantum computer providing reliable outputs to quantum circuits beyond brute force classical computing.
One key question brought forward by this quantum utility experiment was related to validation pathways for the types of quantum computation that enter into the realm of classically intractable computations.
Shortly after this first utility experiment was published, a flurry of very impressive classical solutions to the problem came out ~\cite{Tindall_2024, Kechedzhi_2024, fastclassicalsimulationevidence, patra2024efficienttensornetworksimulation}, with an interesting degree of disagreement between them, similar in magnitude to the very confidence intervals obtained in the utility experiment.
As quantum processors continue to become more capable and less noisy, the question of whether quantum computing can actually serve as the arbiter of these types of problems becomes relevant.
Moving forward, pushing problem (circuit) complexity further, it is difficult to envisage competitive classical solutions to problems involving highly entangled quantum systems.

\begin{figure*}[!t]
\centering
\includegraphics[width=\textwidth]{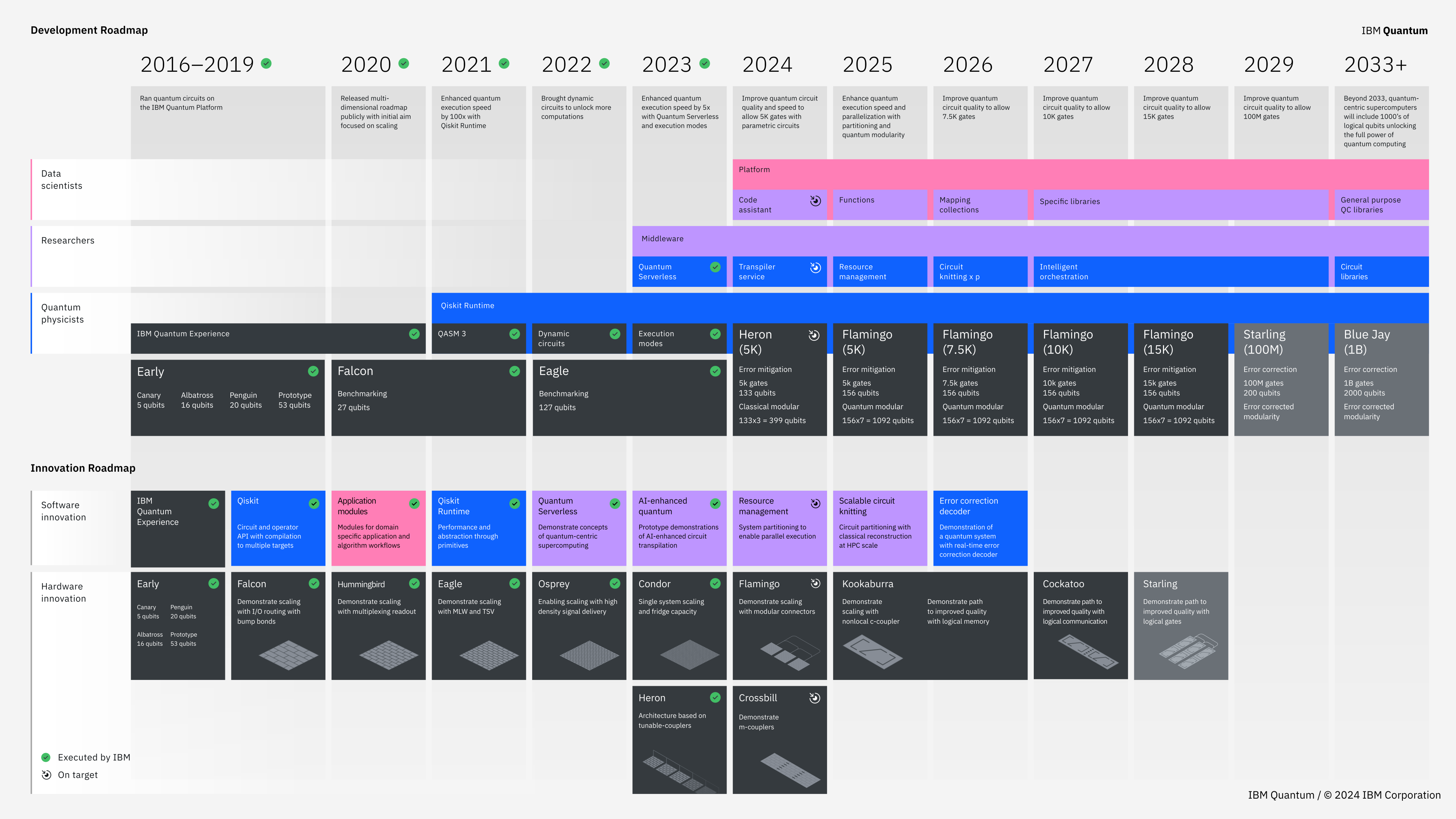}
\caption{IBM Quantum Development and Innovation Roadmaps.}
\label{fig:roadmap}
\end{figure*}

On IBM Quantum’s development roadmap, shown in Fig.~\ref{fig:roadmap}, quantum error mitigation (QEM) will continue to play a major role in quantum computations for the next 5-6 years.
Such methods have proven impactful for reducing the effect of noise inherent in quantum systems.
The roadmap also focuses on enablement, lowering the technical bar for users to begin quantum programming, providing domain-specific libraries, and providing a reliable platform for performing quantum computations, and innovations for modular scaling of systems.

In the next several years, discrete QPU qubit-count will increase modestly while interconnecting multiple QPUs -- giving long-range, non-local connectivity -- via classical and, eventually, quantum communications channels will be pursued.
This ultimately will lead to wider circuits (i.e., larger number of qubits), with superconducting qubit technology improvements providing gains in depth (i.e., coherence times) for executing circuits with $O(10^9)$ gates, including QEC.

\section{Quantum-centric Supercomputing in a Nutshell}
\begin{figure*}[!t]
\centering
\includegraphics[width=\textwidth]{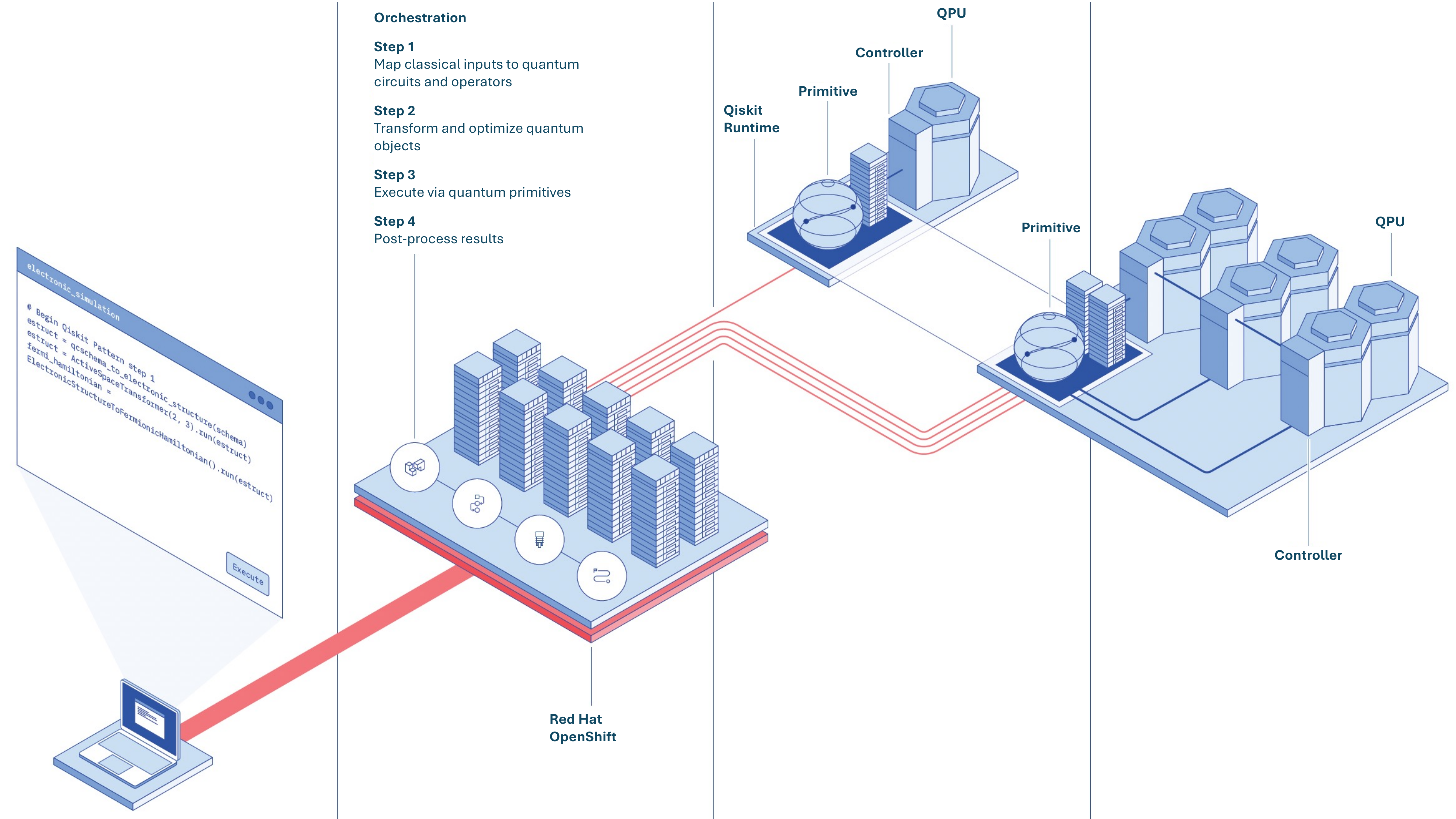}
\caption{Illustration of quantum-centric supercomputer (QCSC) architecture.}
\label{fig:qcsc-arch}
\end{figure*}

So, what is QCSC?
Well, quantum computers will not perform every single computation -- they are best suited for solving specific types of problems, typically those where the data presents correlations that are too complex to tackle with classical means alone.
As such, classical and quantum resources need to work in concert:
that is, classical (including HPC) complementing quantum in utility-scale experiments and beyond.
Figure~\ref{fig:qcsc-arch} illustrates a generic architecture for QCSC.

We have been working closely with domain scientists to learn their needs for quantum.
Several working groups have been formed -- including optimization~\cite{abbas2023quantumoptimizationpotentialchallenges}, health sciences~\cite{basu2023quantumenabledcellcentrictherapeutics}, materials science~\cite{ALEXEEV2024666}, and HEP~\cite{PRXQuantum.5.037001} -- that bring together researchers across cross-cutting disciplines to provide useful quantum computing to these and other domain sciences. 
Points of focus include classical-quantum workload integration and their workflows, programming models to support programmability of QCSC systems, and use cases (classically intractable problems that need to be solved).

A major milestone, of course, is demonstrating practical quantum advantage.
This comprises two main requirements: (1) being able to execute quantum circuits faster on quantum hardware than on classical hardware (this is essentially the definition of quantum utility); and (2) mapping interesting/difficult problems to quantum circuits.
This in essence reduces to the requirement that in order to claim quantum advantage in a practical way, the best way to execute the computation is through a quantum circuit.
Indeed, (1) is not valuable if there exist classical algorithms that bring the solution faster than executing a quantum circuit, or family of quantum circuits.
This mapping can only be done (or more easily be done) with a range of expertise.

The remainder of this document will briefly (here, due to space constraints, and during the talk, due to time constraints) discuss generally topics we believe are important to QCSC:
workload management, hybrid workflows, and programmability in the context of QCSC.
Lastly, we will give a use case which, while targeted, we believe will resonate with the wider audience.

\section{Focus Areas\label{sec:foci}}
We start with workload management.
As already mentioned, QCSC comprises the interplay between classical and quantum resources.
One can consider the workload management layer as a conventional software system, e.g., SLURM, with extensions for interfacing and adapting requirements for quantum devices.
For instance, managing disparate latencies between resources and resource sharing.
Near-term, quantum resources will be limited (i.e., a single or small number of devices), so a focal point is how to efficiently share these resources among users and their applications.
These resources can be incorporated as virtual devices within a classical compute node.
Hybrid workloads are spawned from a login node (as convention) where quantum  can be executed either via a low-latency, high-throughput interface or through cloud access to the quantum hardware.

In the context of workflows, one can envisage near-term and far-term types.
Near-term, or ``weakly-coupled'' regime, where latencies smaller than qubit coherence times are not essential;
distributed classical HPC and quantum resources, or co-located systems with `slow' (i.e., classical-quantum communication latencies $\gg$ coherence times) interconnects.
Such workflows would be temporally decoupled: run the classical part(s) here and now, the quantum there and then.
This is the regime in which we find nearly all current use cases live.

In the far-term, or the ``strongly-coupled'' regime, a high-speed bus, providing communication latencies $\ll$ coherence times, can be envisaged.
Workflows in the strongly-coupled regime would have requirements including and beyond those of weakly-coupled regime, such as single large-scale hybrid applications where the quantum resources are more akin to classical accelerators (i.e., \textit{quantum coprocessors}) available for offloading specific parts of problems in close to real-time.
Although this feels like a natural path for QCSC, there are currently very few instances where such coupling is considered a necessity.
Programmability in the loosely-coupled regime is rather trivial:
classical and quantum tasks are temporally decoupled, can be written in whatever language.
Effectively, the workflows comprise a ``bag of tasks.''
With strongly-coupled systems, this becomes interesting:
one desires to offload quantum-accelerated kernels while classical tasks execute asynchronously.
Here, one can have a single-source hybrid application that pervades all available resources, part of a larger workflow.
A compiled, HPC-like language would be advantageous.

\section{A use case for high energy physics\label{sec:usecase}}
Lastly, we introduce a use case.
This is a bit biased, given the speaker's background, but believe it applies to many folks in the audience.

We consider a typical HEP data production chain comprising two paths:
(1) data, where the input is collected, at least in this generation, classically, some online data processing occurs, followed by processing, and out comes data for analysis;
and (2) the MC path where scattering cross sections (matrix elements) are computed via MadGraph~\cite{Alwall_2014} or some other generator, simulation of the (classical) detector is performed, with the remaining parts (i.e., reconstruction and beyond) converging with the data path.
Where can QCSC fit in here?

One can foresee event generation, manifestly quantum in nature, as a particularly important place where quantum computing can be useful.
Many HEP quantum research has been focused on this area;
rightfully so, as this is where one would expect quantum advantage:
simulating quantum nature.
Recent work includes fermion scattering on a quantum device~\cite{chai2024fermionicwavepacketscattering}, quantum simulation of hadron dynamics~\cite{farrell2024scalablecircuitspreparingground}, and quantum simulation of SU$(3)$ dynamics~\cite{ciavarella2024quantumsimulationsu3lattice}.

Detector simulation is also a highly active area of research, primarily due to the excitement around quantum machine learning (QML).
A typical QML example is to use a promote a generative adversarial network (GAN) to a quantum GAN, where a parameterized quantum circuit is used in conjunction with a classical discriminator~\cite{rehm2024quantumgenerativeadversarialnetwork}.
The classical part of such algorithms is used to calculate gradients and update parameters for subsequent executions of the quantum generator.

Reconstruction-type problems are also seeing some interest.
In particular, tracking detectors, where problems blow up combinatorially with the number of particles involved.
Work from Nicotra et al.~\cite{nicotra2023quantumalgorithmtrackreconstruction} has demonstrated at small-scales, $O(10)$ qubits, that the use of quantum computing can achieve as good or better results for determining particle tracks of interest.
One major issue: depth.
A simple example of a tracking detector comprising three layers, five particles, and 50 doublets required 14 qubits and $O(10^6)$ entangling gate layers.
This is not exclusive to this type of problem but is in general prevalent in HEP.
Seems we have some work to do before realizing this on real hardware.
On the other hand, the algorithm utilizes the HHL algorithm~\cite{Harrow_2009} which promises exponential improvements over classical solutions to solving linear systems of equations.
Such algorithms can thusly prove to be very useful for future generations of tracking detectors.

Lastly, analysis. Thanks to interest in QML and the likes, this is a rich and exciting area of research.
Representational work includes using quantum support vector machines to perform classification of background and signal for $t\bar{t}H$ events~\cite{Wu_2021}.
It was demonstrated that quantum analogs of conventional machine learning techniques can provide background rejection capabilities similar to classical methods by leveraging the large dimensionality of the quantum Hilbert space in favor of the classical feature space.

We note that a possible advantage for QML is comparable performance to classical ML with small fractions of data, though this has not been proven.
It is believed that manifestly quantum problems will be the most applicable to quantum computing, at least in the near term.
Mapping classical problems to quantum is a difficult outstanding question.

\section{Summary of the summary\label{sec:summary}}
We are now in the era of utility, where quantum computing can provide reliable solutions at-par or beyond brute force classical computing methods.
We see QCSC as a system leveraging quantum and classical computing devices to enable execution of hybrid workloads at utility-scale that will assist in providing better quantum results in noisy quantum devices.
Working with partners, clients, and collaborators, we look to domain scientists to find the hard problems to solve, and cross-cutting expertise being required to work on a solution.
We discussed several focus areas for QCSC, including workload management, hybrid workflows, and programmability, with a use case centered around HEP, where it can be envisaged QCSC will play a pivotal role and ultimately demonstrate quantum advantage.

\section{Q\&A\label{sec:qa}}
Note the following Q\&A are not verbatim as the session was not recorded.
We did our best to recollect the questions as posed and the answers as given.

\begin{enumerate}
    \item \textit{Where in the HEP data processing chain do you think QC will be most important?}
\end{enumerate}

I believe those problems that are inherently quantum mechanical will be best suited, at least in the near-term, for offloading to quantum devices.
Mapping/encoding classical problems to is not as natural.
This mapping can also be seen as forcibly done, like trying to get one of Cinderella's evil step sisters' feet into one of her glass slippers.
Sorry, maybe not the best example for the audience.
I have a lot of kids.
The real quantum part [of the HEP data processing chain] is the generation, or hard-scattering processes, and parton showering.

However, looking forward to future experiments wherein detectors employ quantum sensors (A. Chou et al., 2023), quantum computing can play a major role in detector simulations -- you are now potentially dealing with quantum information, and that information can be processed using a quantum information processor.
Having worked nearly a decade on classical detector simulation, this excites me to think about.

\begin{enumerate}[resume]
    \item \textit{(Follow-up) How do you see analyses being performed in next-generation experiments?
    How would quantum sensors improve our analyses beyond providing more precision?}
\end{enumerate}

That is an excellent question, and to be honest, I'm not exactly sure.
The point of quantum sensors is to improve spatial and temporal dimensions of measurements.
With detectors having quantum sensing elements, one can envisage quantum algorithms playing a significant role in object reconstruction and even data analysis if the sensors give us quantum data to work with.
And that would be very interesting.

\begin{enumerate}[resume]
    \item \textit{What is the difference between QEM and QEC?}
\end{enumerate}

In short, QEM are techniques to reduce the impact that noise in current quantum computing systems has on a computation -- basically, it attempts to recover the true answer.
However, it doesn't completely fix errors that occur, rather, it alleviates them.
As of now, QEC imposes serious overheads to actually correct errors either as they occur or at the end of a computation.
It is because of the resource requirements for QEC that we will focus on QEM in the next several or more years.

\textit{Addendum}: QEM can go beyond reducing the impact of noise: it can produce noise-free expectation values (with some error bar) of observables. There are two main aspects of differentiation between QEM and QEC in how they enter computations, the way I see it. One is that QEM is a non-deterministic approach. This essentially means it is executed on ensembles of outcomes. In contrast, QEC works at a single shot level. And the second is that QEM has an unfavorable exponential overhead which depends on noise and problem size, whereas the overhead of QEC is polynomial (and for single-shot versions of QEC it can be linear, I believe).

\begin{enumerate}[resume]
    \item \textit{Can you explain further what a ``virtual QPU'' is?}
\end{enumerate}

It can be thought of as a placeholder for a real piece of hardware.
That is, the node shown with a `vQPU' in the slide does not have a physical QPU, or quantum device, connected to it.
Since quantum resources will be limited in the early years of QCSC, they will necessarily be shared among users;
cloud providers do this now with classical accelerator hardware.
For today's use cases, a QPU does not need to always be available or even have a persistent connection to classical HPC.
In short, the `vQPU' is a shared artifact that the workload management system uses to provide the resource when needed or is available.

\bibliography{main}
\end{document}